\documentclass{birkart04}
 \theoremstyle{definition}
 
 \theoremstyle{remark}

 \numberwithin{equation}{section}

\begin{document}
%--------------------------------------------------------------------------
% editorial commands: to be inserted by the editorial office
%
%\firstpage{1}
%\volume{228}
%\DOI{003-0001}
%\seriesextraline{This is the Concrete Title of this Book\br H.E. Rowe and S.T.C. Wore, Eds.}
%\commby{inhouse GRRR}

\title[Lattice gas models and KMC simulations]
 {Lattice gas models and Kinetic Monte Carlo simulations of epitaxial
  growth}
%----------Author 1
\author[Biehl]{Michael Biehl}
\address{%
University Groningen, Institute for Mathematics and Computing Science  \\
P.O. Box 800, 9700 AV Groningen, The Netherlands}
\email{biehl@cs.rug.nl}

%\subjclass{Primary 99Z99; Secondary 00A00}

%\keywords{Class file, journal}

% \date{January 1, 2004}
%\dedicatory{To my parents}
%%% ----------------------------------------------------------------------

\begin{abstract}
 A brief introduction is given to Kinetic Monte Carlo (KMC) simulations
 of epitaxial crystal growth.
 Molecular Beam Epitaxy (MBE) serves as the
 prototype example for growth far from equilibrium. 
 However, many of  the aspects discussed here would carry 
 over to other techniques as well. 
 A variety of approaches to the modeling and simulation of  
 epitaxial growth have been applied. 
 They range from the detailed quantum mechanics treatment of  
 microscopic processes to the
 coarse grained description in terms of stochastic 
 differential equations or other continuum approaches. 
 Here, the focus is on discrete representations
 such as lattice gas and Solid-On-Solid (SOS) models.
 The basic ideas of the corresponding  Kinetic Monte Carlo 
 methods are presented. Strengths and weaknesses become apparent
 in the discussion of several levels of  simplification 
 that are possible in this context. 
\end{abstract}

%%% ----------------------------------------------------------------------
\maketitle
%%% ----------------------------------------------------------------------
%\tableofcontents

\section{Introduction} \label{introduction}

An impressing variety of methods is applied in the
theory, modeling, and simulation of epitaxial growth.
 They range from the faithful
 quantum mechanics treatment of the microscopic dynamics to the
 description in terms of coarse grained continuum models. 
An overview of the field can be obtained from, e.g.\ 
\cite{markov,barabasy,villain,krug} and this volume.

The purpose of this contribution is to provide a brief
introduction to one of the most  widely used approaches: 
the Kinetic Monte Carlo (KMC) simulation of lattice gas models. 
It is not intended to give a detailed exhaustive review or historical summary 
of the many aspects of this line of research.
The aim is to introduce and illustrate some of
the basic concepts of the method and to provide a starting point
for the interested reader. 
The selection of examples from the literature 
also reflects this purpose  and the list of references
is far from being complete in any sense. It is the result of
a very personal and often random choice of examples.
Numerous important contributions to the advancement
of the KMC technique and its applications cannot be mentioned
here  and  the interested reader is directed to, e.g.\ 
\cite{markov,barabasy,villain,krug,newman}
for further references. 

Among the different realizations of epitaxial growth,
Molecular Beam Epitaxy (MBE) is a particularly clear-cut one
\cite{markov,barabasy,villain,krug}.
In this technique, one or several adsorbate materials are heated in an oven
which is contained in an ultra-high-vacuum chamber. 
The evaporating particles form an atomic or molecular beam which
is directed onto a substrate crystal. Arriving particles
are incorporated and contribute to the growing film upon the substrate. 
The term homoepitaxy is used if the deposited adsorbate  and
the substrate material are identical, whereas in heteroepitaxy
they differ. 
Apart from the selected materials, the most important 
experimental control parameters in MBE growth  
are the substrate temperature and the flux of incoming
particles. 

    \begin{figure}[t]
    \begin{center}
     \includegraphics[scale=0.40]{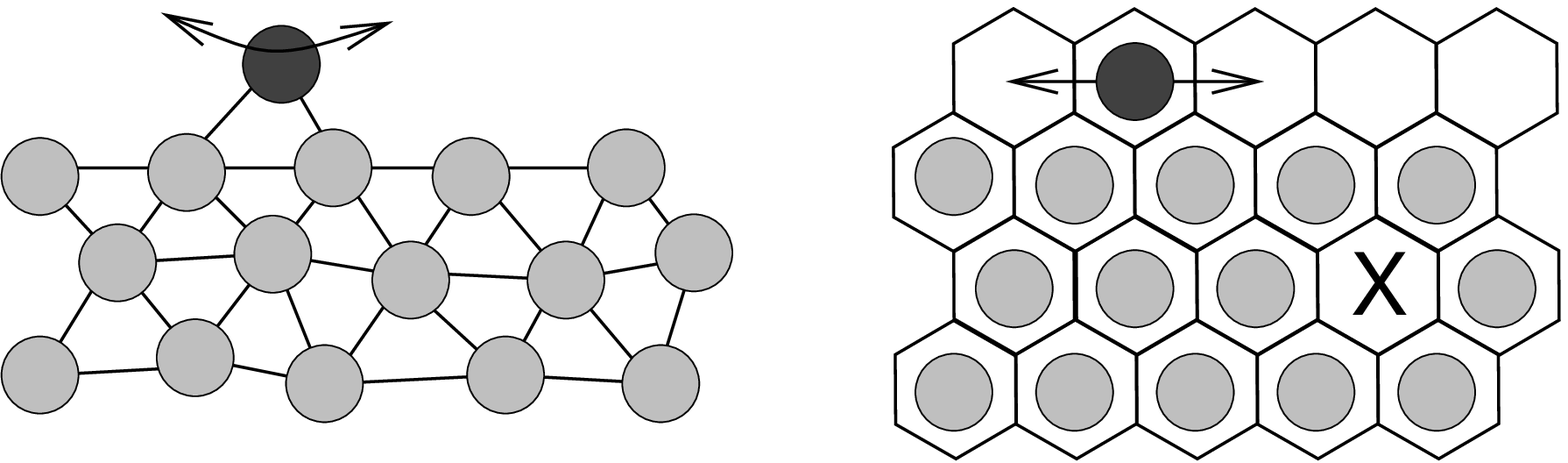}
     \caption{\label{illustration} 
     \protect{\newline}
     Cartoon of hopping diffusion on a flat surface (cross section).
     \protect{\newline} 
     Left: Particles can assume continuous positions in space, straight lines
     correspond to nearest neighbor bonds. In an MD simulation,
     thermal fluctuations of the entire system may result in a jump 
     of the adatom (dark) to a neighboring
     pair of surface atoms.   
     \protect{\newline} 
     Right: Simplifying lattice gas representation of the same situation.
     Atoms can only occupy pre-defined lattice sites as represented by
     the hexagonal cells, here. In a KMC simulation, the 
     adatom jumps to the left or right neighbor cell with a given rate.
     The cross marks a vacancy in the bulk  
     which would be, like overhangs, {\sl forbidden} in Solid-On-Solid
     models. 
     }
     \end{center}
     \end{figure}

MBE has become a well-established technique for the 
production of high-quality crystals, as it
allows for a very precise control of the
growth conditions. It is, for instance, possible 
to add  monoatomic layers of a compound semiconductor
to the growing film by alternating
deposition of the elements in  Atomic Layer Epitaxy (ALE).

Molecular Beam Epitaxy is applied, for example,
in the growth of  
layered semiconductor heterostructures 
for electronic devices or in the development of 
thin magnetic films for novel storage media. MBE plays also a significant role
as a tool in the design of nano-structures, such
as Quantum Wires or Dots \cite{QD}. 
The range of  materials used in MBE includes
conventional semiconductors, elementary metals, metal-oxides,
and organic molecules. 

Growth in an MBE environment is clearly far from equilibrium, as the system
is constantly driven by the deposition flux and an extremely
low pressure is maintained in the vacuum chamber. This is different
from, say, growth from the vapor which can be much
closer to thermal equilibrium. 
This feature makes MBE highly attractive from a theoretical point
of view. It provides a workshop in which to 
put forward analytical approaches and develop  tools for 
the simulation of more general  non-equilibrium systems.

   \begin{figure}[t]
   \begin{center}
     \includegraphics[scale=0.40]{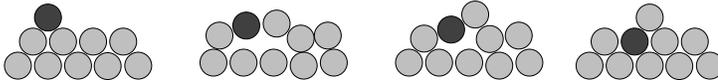}
     \caption{\label{austausch}
     Illustration of an exchange diffusion process (cross section).
     A concerted move of several particle leads to
     the exchange  of the adatom (dark) with a surface atom. 
     Note that, if all atoms represent  the same chemical element,
     the final states of hopping and exchange diffusion in this
     illustration would be identical whereas  the transition states 
     differ significantly.  }
   \end{center}    
   \end{figure}

In the following section, relevant atomistic processes and
some approaches to the theory and modeling of MBE growth 
are discussed.  Many aspects would directly carry over
to the physics and modeling of similar growth techniques. 
In section \ref{kmc}, the basic features of lattice gas models 
and the corresponding Monte Carlo kinetics are presented
and a short conclusion is given in
section \ref{summary}. 

\section{Atomic scale processes in MBE growth} \label{levels}

In MBE, physical processes in a very wide range of time  
and length scales are relevant. Accordingly, a set of quite different
methods of modeling is required when aiming at a 
more or less complete
 realistic picture of the growth process. 

Ultimately, all macroscopic features  of the growing surface 
emerge from the interactions of atoms and their microscopic
kinetics. Therefore, a faithful quantum mechanical 
description  on the atomistic level is clearly desirable.
The so-called {\sl Density Functional Theory\/} (DFT)  is particularly
suitable in this context \cite{dft}. This method is based on the,
in principle exact, description of many electron systems in terms
of the electron density only. Despite the clear advantage 
over an explicit treatment of the many particle Schroedinger
equation, the computational cost of the method is significant. 
It is therefore often applied to relatively small systems of
a few atoms or molecules, or to unit cells of periodic structures.
\footnote{The use of relatively vague terms like {\sl relatively small
systems\/} relates to the fact that their meaning steadily changes 
with the rapid evolution of available computer power.}

As just one example for the latter, by means of DFT calculations
it is possible to evaluate  
and compare the specific ground state energies of different 
reconstructions of a particular material surface. 
The calculation yields also the precise 
structure of the reconstruction, i.e. the location of the atoms.
Clearly such information is valuable in the development
of models describing the surface or its evolution in a
non-equilibrium situation.  For examples in the context
of compound semiconductor surfaces see \cite{itoh,prbtwosix}
and references therein.  

The analysis of dynamical properties such as diffusion
processes is obviously more involved than 
the investigation of ground state properties only
\cite{kratzer}.
Impressing progress has been achieved in recent years 
in the combination of the DFT approach with other methods
such as Kinetic Monte Carlo simulations. Reference \cite{kratzer}
gives an introduction to this promising line of research.  

Often, the results of DFT calculations
depend crucially on the quality of
the actual approximations made, which can
lead to significant discrepancies in the literature. 
As an example, a discussion and comparison
of energy barriers obtained for diffusion on 
various fcc(111) metal surfaces is given in  \cite{krug}. 

One of the most important tools for describing the 
temporal evolution of many particle systems is
Molecular Dynamics (MD), see for instance
\cite{md} and the contribution by K. Albe to
this volume \cite{albe}. In its simplest 
version it amounts to the numerical integration of 
Newton's equations of motion based
on classical interactions of the particles.
Methods have been devised that allow for
imposing a constant temperature or other
physical conditions on the system \cite{md}. 

Frequently, the interactions are given in terms
of classical pair or many particle potentials. 
These can range from simple model interactions to 
highly sophisticated material specific potentials.
The latter can be obtained from first principles, e.g.
by means of DFT calculations,  or result from fitting
suitable parameterizations to experimentally observed 
material properties. 
It is also possible to incorporate the  quantum mechanics
of the system directly into the MD scheme, as for instance
in the celebrated DFT-based Car-Parrinello method \cite{parrinello}.

Efficient implementations and modern computers 
allow for the simulation of relatively large systems. 
However, a serious restriction of all MD techniques remains: the
physical real time intervals that can be addressed are
usually quite small, see the discussion in 
\cite{albe} (this volume).
Even with sophisticated acceleration techniques, such as
proposed in \cite{voter}, it seems currently 
infeasible to reach MBE relevant time scales of seconds
or minutes.  
MD aims at a faithful representation of the microscopic dynamics
in continuous time. The relevant time scale of atomic vibrations
in a solid is, for instance, $10^{-12}s$. Hence, an MD 
procedure should advance time by even smaller steps  in the
numerical integration. 

Consider, as an example, a single
adatom on top of a flat surface as sketched in Figure
\ref{illustration}. Thermal fluctuations may result in 
a {\sl jump} of the adatom to a neighboring pair of surface 
atoms.  However, such
an event  will occur quite rarely.  
A large portion of an MD computation will be used to simulate 
the collective vibration of the crystal which does not
change the system significantly over relatively long
periods of time. 
It is precisely this difficulty which motivates the
basic idea of KMC simulations, see section
\ref{kmc}.

    \begin{center}
    \begin{figure}[t]
     \includegraphics[scale=0.45]{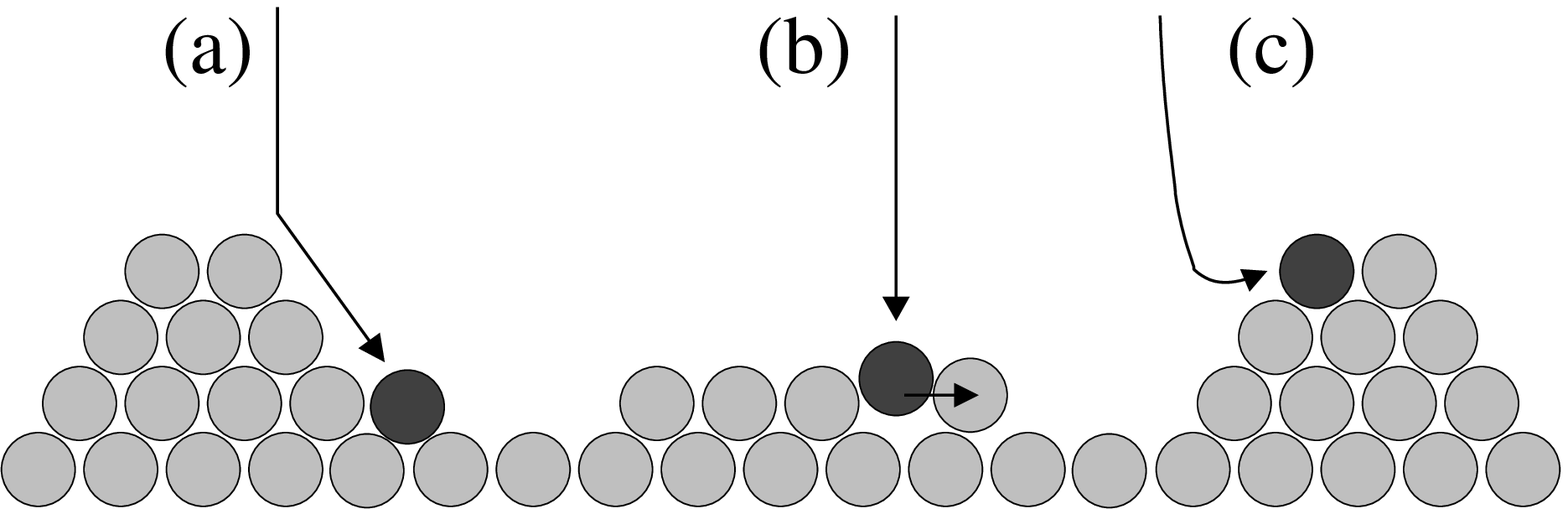}
     \caption{\label{landing} 
      Some kinetic effects that may play
      a role in the incorporation of deposited adatoms. 
      \protect{\newline}
      (a) {\it Downhill funneling}: the deposited particle
      slides down a slope until a local minimum of
      the surface height is reached. 
      \protect{\newline}
      (b) {\it Knockout process}: the momentum of the arriving
      particle suffices to push out a surface adatom at a
      terrace edge. 
      \protect{\newline}
      (c) {\it Steering effect}: attractive forces can 
      influence the trajectory of the arriving particle
      significantly. Obviously, (a) and (c) are unlikely to occur
      both under the same conditions. }
    \end{figure}
    \end{center}

As the discussion in section \ref{kmc} will show,
perhaps the most attractive advantage of Molecular Dynamics
over standard KMC techniques
is that no {\sl a priori} assumptions about the 
possible microscopic processes have to be made.  
As an example, it is not necessary to know in advance whether 
adatom diffusion occurs by {\sl hopping\/} or by 
concerted {\sl exchange processes\/}  involving many particles,
see Figure \ref{austausch} for an illustration.
A similar problem is that of island mobility: in several materials
it has been observed that, after mobile adatoms meet and coalesce,
they may perform  diffusion as a more or less rigid cluster on the surface.
Such cluster diffusion can influence the growth kinetics significantly \cite{albe}.
In principle, an MD simulation which is based on a faithful 
representation of atomic interactions will produce all relevant 
single particle or concerted moves {\sl automatically\/} \cite{albe,voter}.
In contrast, one usually has to pre-define
a catalogue of possible events
in KMC simulations, see section \ref{kmc}.

Molecular Dynamics should also be the natural technique for 
simulating processes which cannot be described as thermally activated.
This concerns in particular the {\sl transient mobility\/} 
of freshly deposited adatoms on the surface. 
A more detailed discussion of such processes can 
be found in \cite{krug}, which also provides references
to related experimental studies of specific materials. 

A deposited particle may arrive with a certain momentum
at the surface. Not only was the particle part of a
directed beam, it may also pick up additional kinetic energy
when it is attracted by the surface.  
The term {\it downhill funneling\/} describes one
possible consequence:
an adatom that arrives at an inclined portion
of the surface may slide down the slope until
it reaches a local minimum in surface height, see
Figure \ref{landing} (a) for an illustration of the process.
It is clearly to be distinguished from thermally 
activated diffusion which occurs only after the 
particle has been incorporated into the crystal surface.

Another possible effect is  sketched in Figure \ref{landing} (b):
the arriving particle may  {\sl knock out\/} an atom at a terrace
edge. This process will smoothen a surface and would favor
layer-by-layer growth, effectively.

The interaction of the deposited atom with the surface may have
a significant impact on the particle's trajectory and may lead to
a deviation of the adatom position from the deposition site, 
cf. Figure \ref{landing} (c). 
Such  {\sl steering effects\/} have been observed explicitly
in, e.g., metal epitaxy, see also \cite{krug} for references. 

The  capture of arriving adatoms in weakly bound intermediate states 
may play an important role as well. For several compound semiconductors 
it is assumed that
one of the elements may reside in a highly mobile state close
to the surface before a regular lattice position is reached. 
Examples in the literature discuss the role of
mobile As in GaAs epitaxy \cite{itoh} or Te in ALE growth of CdTe
\cite{prbtwosix}. 

Frequently, the deposition process is complicated by the
fact that molecules or small clusters of atoms, e.g. dimers,
arrive at the surface. Their dissociation may be a highly complex
process which is not governed by thermal activation.
Again, MD should be the method of choice for the
simulation and understanding of the deposition kinetics.
In lattice gas models and KMC simulations 
such processes can be implemented in a highly simplifying effective
fashion, at best. 

In spite of its practical restrictions, Molecular Dynamics has
been and continues to be extremely useful in the context of epitaxial 
growth, see \cite{albe} for specific examples.
Simplifying faster techniques like Kinetic Monte Carlo simulations rely 
in general on the {\sl a priori\/} knowledge of the catalogue of relevant
microscopic processes. In particular when aiming at material specific
investigations, MD can provide valuable detailed insights in this context.

\section{Kinetic Monte Carlo simulations and lattice gas models}  \label{kmc}

\subsection{Thermally activated processes\/} \label{activated}
The Kinetic Monte Carlo method 
extends the range of accessible time scales by 
implementing appropriate {\sl short cuts\/} in
the treatment of thermally activated processes.  Only those configurations
are explicitely taken into account, in which the system resides
for significant time intervals. A stochastic process is defined
which takes the system from one such state to a neighboring
one in configuration space.  
Obviously, the physical properties
must be taken care of when specifying the transition
rates.

\begin{figure}[t]
    \begin{center}
     \includegraphics[scale=0.30]{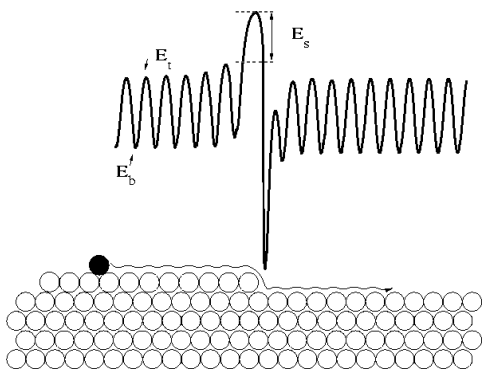}
     \caption{\label{pes} 
      Sketch of the potential energy surface (PES) for
      single adatom hopping diffusion on a terraced surface, 
      courtesy of Florian Much \cite{muchdiss}.
      In this particular example, pairwise interactions 
      according to a Lennard-Jones potential were assumed, see
      \cite{muchdiss} for details.  $E_b$ is the energy of
     a binding state, whereas $E_t$ corresponds to the
     transition state of hopping diffusion. The Ehrlich-Schwoebel
     barrier $E_S$ hinders inter-layer hops at the terrace
     edge.
     }
    \end{center}
    \end{figure}

In order to illustrate the basic idea of the KMC approach,
we return to the above example of 
hopping diffusion, c.f. Figure \ref{illustration}.
A simplifying interpretation of such a situation is illustrated
in Figure \ref{pes}: if we assume the rest of the crystal is
{\sl frozen\/}, the adatom moves in a potential energy surface 
(PES)  that results from its interactions with the other atoms.
In reality, the underlying crystal rearranges and reacts
to any movement of the adatom. However, as long as this does not
lead to topological changes, the {\sl frozen crystal picture\/}
is essentially valid and appropriate for the following considerations,
see also \cite{offl} (this volume).
The  local minima in the PES, cf. Figure \ref{pes}, 
correspond to the relevant configurations in the KMC approach
and are termed the \textit{binding states}.
As discussed above, thermal fluctuations will occasionally 
provoke a jump from one such minimum to a neighboring one.
One implicitly assumes that the time required for the 
actual transition can be neglected in comparison with the
time spent in the binding states. 
Under further simplifying assumptions, only some of which will
be discussed below,
the typical waiting time for  a transition  is given by
an Arrhenius law of the form
\begin{equation}  \label{arrhenius}
 \tau  \, =  \, \tau_o \,  \exp\left[ \frac{(E_t - E_b)}{k_B T} \right]
 \qquad \mbox{or} \qquad
 r  \, =  \, \nu_o \,  \exp\left[ -\frac{(E_t - E_b)}{k_B T} \right]
\end{equation}
for the corresponding rate, respectively, where
$k_B$ is Boltzmann's constant. The rate $r$ decreases exponentially with the height 
$(E_t-E_b)$ of the energy barrier
that has to be overcome in the process , cf. Figure \ref{pes}.
$E_b$ is the energy
of the so-called {\sl binding state\/}, whereas $E_t$ corresponds
to the {\sl transition state}.  
The barrier is  compared with the typical thermal energy
$k_B T$ in the system, i.e.\ the higher the  temperature $T$ the more
frequent becomes the event. 
The prefactor $\nu_o$ is termed the {\sl attempt frequency}.
In the Arrhenius law it is assumed to be independent of the temperature.

Figure \ref{pes} displays the PES as for a single particle
in the vicinity of a terrace edge. In the center of the 
upper or lower terrace, the potential is oscillatory with 
equivalent minima at the regular lattice sites and identical 
barriers. Near the  edge, however, the shape of the PES shows  some
distinct features: a very deep 
minimum is found right at the step
due to the good coordination, i.e. the interaction with many neighbors. 
Consequently, a particle that has attached to the upper terrace
will detach with a relatively low rate only. A second pronounced
feature is the additional, so-called Ehrlich-Schwoebel barrier 
$E_S$  \cite{markov,barabasy,villain,krug} for hops from the upper
to the lower terrace (or vice versa). Here, the transition state
is very weakly bound and hence {\sl interlayer diffusion} is hindered
effectively. This can result in the formation of pronounced mounds 
 in the course of growth, see \cite{barabasy,villain,krug}
 for discussions of the effect.

In Transition State Theory,  the Arrhenius law is  motivated
by arguing that the occupation of binding  and 
transition states should correspond to an effective thermal equilibrium situation.
Following this line of thought one obtains directly the form of Eq.\ (\ref{arrhenius}).
This and other approaches to the evaluation of transition rates
are discussed briefly in \cite{krug,kratzer,voter} and
in greater detail in e.g.\ \cite{chemistry,haenggi}. 

In principle it is also possible to determine the attempt frequency by
taking into account the (vibrational) entropies, i.e. the free energies
of binding and transition state, see \cite{krug,kratzer,chemistry,haenggi}
 and references therein.
However such an evaluation can be quite involved and
has to go beyond the frozen crystal approximation. Consequently, good
estimates of attempt frequencies are hardly available in the literature. 
Very often it is therefore simply assumed that the attempt frequency is
the same for all possible diffusion processes in the system.
A popular choice is $\nu_o = 10^{12}/s$, corresponding to the typical
frequency of atomic vibrations.    
Usually, there is no obvious justification for this crude 
simplification. In fact, several authors have argued that it is 
simply incorrect in particular cases, see \cite{krug} for example 
references. 

From a practical point of view, the simple picture of
a single attempt frequency has several striking advantages:
with only one $\nu_o$ in the system this quantity 
can be taken to define the unit of time 
and does not have to be taken care of explicitly. 
More importantly, Arrhenius rates with identical attempt frequency
are guaranteed to satisfy the condition of detailed 
balance \cite{krug,newman}: for two
neighboring binding states $b_1$ and $b_2$  with energies
 $E_1,E_2$ one obtains directly
 \begin{equation} \label{detailedbalance}
 \frac{ r(b_1 \to b_2) }{r(b_2 \to b_1)} 
 \, = \, \frac{ e^{-E_2/(k_B T)}}{e^{-E_1/(k_B T)}} 
 \end{equation} 
 as $\nu_o$ and the dependence on $E_t$ cancel
 in the ratio. 
 The detailed balance condition is by no means a necessary requirement 
 for meaningful simulations, especially not in situations
 far from equilibrium.
 However, it conveniently guarantees that the system --
 in absence of deposition and desorption -- 
 would approach the correct thermal equilibrium
 eventually, i.e.\ a 
 Boltzmann distribution of binding states \cite{newman}.
 If it is violated one might have to worry, for instance,
 about unphysical cycles of diffusion events in which the system 
 could pick up (free) energy constantly \cite{krugremark}.

 Note that deposition and desorption in MBE models necessarily
 violate detailed balance. Deposition is not an activated 
 process anyway and the discussion of detailed balance is meaningless,
 the same is true for the kinetic processes illustrated in
 Figure \ref{landing}.
 In desorption a particle
 has to overcome a barrier which is simply identical with
 its binding energy at the surface. Hence, desorption can be
 implemented with the corresponding Arrhenius rate. However,
 the process is irreversible: in an ideal MBE environment a 
 perfect vacuum is maintained in the chamber and desorbed
 particles will be removed immediately. The situation differs
 significantly from a surface that may exchange particles
 with a surrounding vapor or melt.

The validity of the Arrhenius law for thermally
activated processes, Eq.\ (\ref{arrhenius}),
hinges on several assumptions, which will not
be discussed in detail, here. 
A more thorough discussion of these
conditions and further references
can be found in \cite{krug,voter,chemistry,haenggi}, for instance. 
One important requirement is, loosely speaking,
that the minima in the energy surface 
have to be deep enough to guarantee that \\
(a) the system remains in the binding state for a significant time interval --
  Arrhenius rates describe only {\sl infrequent events} correctly.\\
(b) once the transition state is reached, the system 
   is caught in the neighboring minimum, indeed.  \\
The importance of these conditions can be demonstrated by a counter-example:
if an adatom moves in a PES with very shallow minima, thermal fluctuations
might very well provoke double or multiple jumps directly to
more distant binding states. According to \cite{doublejumps} the rate for
double jumps can still be written in a form similar to 
Eq.\ (\ref{arrhenius}) but with an explicit temperature dependence of
the  prefactor: $\nu_o \propto \sqrt{T}$.

Most KMC simulations in the literature 
are indeed based on the above simple Arrhenius picture. 
In fact, further simplifications are frequently applied,
some of which will be discussed in the following section.

\subsection{Lattice gas and Solid-On-Solid models} \label{latticegas}

 In principle it is well possible to implement the above
 basic ideas of KMC in off-lattice models 
 with continuous particle positions,
 see \cite{offl} for an example in this volume. 
 It requires the evaluation of the relevant barriers from the PES
 {\sl on the fly\/}.  
 The vast majority of  KMC studies, however, applies important
 additional simplifications. 

 If the material is expected to crystallize in a regular lattice
 without dislocations or other defects, one might as well pre-define
 the set of potential lattice sites.
 In such lattice gas models, a particular site is either
 occupied or empty and particles can only be placed precisely at one
 of the sites as illustrated in  Figure \ref{illustration}.
 Diffusion processes are then represented by hops between lattice
 sites.

\begin{center}
    \begin{figure}[t]
     \mbox{\includegraphics[scale=0.46]{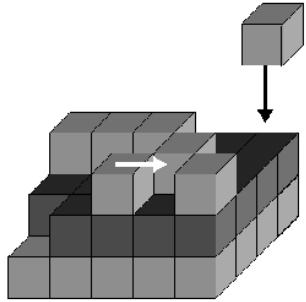}}
     \caption{\label{wuerfel}
     Illustration of a simple cubic Solid-On-Solid lattice gas model.
     For clarity alternating layers are colored differently.
     As overhangs
     and bulk vacancies are excluded, the surface is fully characterized
     by an integer array of height variables above a square lattice substrate.
     As example processes, the deposition of a new particle (black arrow) and
     the diffusion hop of an adatom attaching to neighboring
     particles  (white arrow)  are shown. }
    \end{figure}
    \end{center}

 The popular class of Solid-On-Solid (SOS) models
 fulfills further conditions. Assuming that vacancies in the
 bulk or {\sl overhangs\/} do not occur in the system, one ends
 up with a crystal that is uniquely described by the height of the 
 surface above the substrate.
 A simple cubic model crystal, for instance,
 is then fully specified by an integer array of variables
 above a substrate square lattice, see Figure \ref{wuerfel} for
 an illustration. 

 The layout of the actual lattice structure, the
 design of a catalogue of relevant events,
 and the assignment of rates can be more or less
 closely related to the physical reality. The 
 necessary or adequate level of sophistication clearly depends
 on the problem under investigation. 
 The following paragraphs highlight only a few of the many
 choices one has to make in the design and simulation of 
 lattice gas models.  

 \paragraph{\ \\The lattice\\} 
 In principle, all relevant lattice structures can be implemented
 in a Solid-On-Solid fashion.  When aiming at material
 specific simulations, the correct topology should be
 represented in the model as it determines, for instance,
 neighborhood relations and the potential diffusion events.
 The practical implementation of several
 lattice types is discussed, for instance, in \cite{newman}.
 To name just one specific example: the (001) surface of a
 compound zinc-blende structure is represented as an 
 SOS model in reference \cite{prbtwosix};  the system
 can be treated as the superposition of four simple cubic
 sublattices.  

 One aspect that deserves particular attention is the
 geometry of the substrate and the use of 
 appropriate periodic boundary conditions in order to 
 reduce boundary effects in the simulation of finite systems.
 Imposing an inadequate geometry, e.g.\ using a 
 square shape substrate in the SOS representation of an fcc(111)
 surface, may result in subtle difficulties. For example,
 the shape of large islands that align with the lattice axes  
 might conflict with the substrate geometry. In order to avoid
 such artefacts, the design of the substrate 
 should reflect the lattice 
 symmetries, see \cite{newman} for recipes in particular cases. 

 Monoatomic simple cubic models are employed frequently  in the
 literature.  Whereas they obviously cannot describe any
 real material faithfully, they serve as prototype systems for 
 the investigation of many basic and qualitative features of
 epitaxial growth.  Several properties of growth processes
 are believed to be {\sl universal\/}, i.e. they should not
 depend on the lattice type and other details of the model,
 see e.g. \cite{latticetypes} and references therein. 
 In the investigation of, for example, scaling laws in kinetic
 roughening one resorts to the simplest
  models available \cite{barabasy,villain}.

 \paragraph{\ \\ Deposition and transient mobility \\}
 An ideal beam of particles in MBE growth deposits, on average,
 the same amount of adsorbate material per unit time and
 area everywhere on the substrate. 
 Fluctuations on the atomistic level are
 usually represented in SOS models by choosing
 one of the substrate lattice sites with equal probability
 for each deposition event. 
 Almost always it is assumed in MBE models  that single atoms
 arrive at the crystal, despite the fact that the beam
 often contains molecules or small clusters of a few atoms
 which dissociate at the surface.

    \begin{figure}[t]
    \begin{center}
     \mbox{ \includegraphics[scale=0.32]{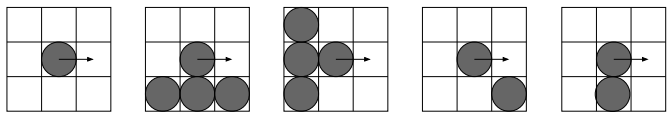}}
     \caption{\label{biham}
     Diffusion hops of an adatom on a crystal surface
     with cubic symmetry, illustration after \cite{biham}.
     In each picture, the  adatom in the center 
     is assumed to hop to the right, and four
     different configurations of the neighborhood
     ($3\times3$ sites) are shown.
      In principle there are
     $2^{7}=128$ possible configurations, due to symmetry only
     $72$ situations are distinct. 
     From left to right, the illustrations correspond to
     diffusion on a flat surface,
     movement along a terrace (step edge diffusion),
     detachment from an edge, and one possibility for the 
     formation of a pair and its dissociation.  }
    \end{center}
    \end{figure}

 Effects of transient mobility upon deposition,
 like the above discussed {\sl downhill funneling} or
 {\sl knockout} at terrace edges, can be represented 
 very conveniently in SOS models.
 A simplifying realization of funneling, for instance,
 amounts to the deterministic search for the lowest
 surface height in a neighborhood of the deposition site. 
 Clearly, such implementations cannot take into account
 the microscopic details faithfully, it is only possible
 to capture the most essential features.   

 \paragraph{\ \\ Adatom mobility\\}
 In the conceptually simplest models, only
 the most recently deposited particle is considered mobile
 at a given time.
 It may, for instance, move to an available empty site of 
 lower height in a neighborhood of the deposition site
 and then become immobile. 
 Thereafter, the simulation proceeds with the deposition
 of another single adatom.   
 Often, only the immediate incorporation of the particle upon arrival 
 at the surface is modeled, whereas activated diffusion over longer
 distances is completely neglected. 
 Clearly, this is computationally very cheap
 and allows for the simulation of large systems and 
 very thick films of, say, $10^6$ layers or more.
 Consequently, such {\sl limited mobility models\/} are employed, 
 for example, in the investigation of basic phenomena like 
 kinetic roughening of  self-affine surfaces, see e.g. \cite{barabasy}.
 Their applicability in material specific modeling is, however, 
 rather limited in general.

 In contrast, {\sl full diffusion models\/} consider all
 atoms mobile at the same physical time, in principle. In the simulation,
 deposition is only one possible event among all other processes
 and has to be implemented with the corresponding rate.  
 As many adatoms move simultaneously the system displays
 a much broader spectrum of possible events.

 In the standard SOS approach only atoms right at the
 surface are considered mobile. Bulk particles are
 surrounded by occupied lattice sites and vacancies
 are excluded. Unless concerted moves or exchange processes
 are implemented explicitly, the entire bulk remains 
 fixed. 
 In the example of the simple cubic lattice this
 could mean that only particles with less then six nearest 
 neighbors are considered mobile at all. 

 In fact, in many cases the adatom mobility is further restricted
 due to simplifying assumptions in the  model design, see 
 the discussion in the next session.
 Some of these are made for practical reasons mainly, others
 reflect  peculiarities of the non-equilibrium growth  
 conditions.

 \paragraph{\ \\ The catalogue of events and rates \\}
 The key step in designing a KMC lattice gas model
 is the setting up of the catalogue of possible events.
 Closely related and equally important is  the
 assignment of rates to the considered processes. 
 
 As an illustrative example we consider the 
 diffusion of a single adatom within one layer of
 a monoatomic crystal surface with cubic symmetry,
 cf. Figure \ref{biham}.
 Already simplifying, let us assume 
 that only hops to nearest neighbor sites occur and
 that their rates depend
 only on the configuration of a very small neighborhood of, say $3\times3$ 
 sites.  Only four examples are shown in the
 illustration, Figure \ref{biham}, in total there are $72$  
 essentially different  neighborhood configurations.
 For each of those a different rate has to be specified 
 in the model, in principle.

 For several material systems, such detailed catalogues of rates
 have indeed been worked out. For the particular example
 of diffusion of a single adatom on an fcc(001) metal surface,
 see for instance \cite{barkema,bihamold,biham}. 
 Note, however, that the evaluation of rates by first principles
 or approximate methods requires a careful analysis of the 
 physical process.  For instance, the rate of hopping and potential
 exchange processes must be compared for each of the configurational
 changes and the task of finding the correct transition path can be
 very demanding, see e.g. \cite{mousseau}. 

 The faithful representation
 of all relevant events in a KMC simulation can be quite involved.
 In our example, inter-layer hops at terrace edges and the like
 have not even been included yet. The potential extension to 
 island mobility and other concerted processes would make the 
 catalogue of events  even more complex.  
 For practical reasons or because 
 the required detailed information about the rates is 
 simply not available, one frequently resorts to simplifying schemes. 
 The aim is to capture the essential features
 of a material system or to study qualitative properties of
 the growth process. 

 In some cases it is possible to find
 efficient parameterizations of the relevant barriers in 
 terms of a small number of independent quantities. 
 {\sl Bond counting schemes\/} have been particularly successful
 in this context \cite{krug,newman,bihamold,schroeder}. 
 The idea is to consider only very few 
 distinct barriers, but to take into account the energies of the
 involved binding states explicitly. In particularly simple
 schemes only the energy of the initial state determines
 the rate, whereas, in general, initial and final 
 configuration are considered. 
 In our example we could
 assume that the quantity $E_o$ plays the role of the characteristic 
 barrier for all nearest neighbor hops within the layer
 and assign a rate of the form
 \begin{equation} \label{bondcountingrate} 
 r \, = \, \nu_o \,  \exp [ -E / (k_BT) ] \qquad
 \mbox{with} \qquad  E = \max \{ E_o, E_o + (E_f - E_i) \}.
 \end{equation}
 Here, $E_i$ and $E_f$ are the energies of the initial and
 final state, respectively. In the simplest cases, their determination
 amounts to counting nearest or next nearest neighbor bonds 
 adding up the associated binding energies. 
 The rate (\ref{bondcountingrate}) is of the  Arrhenius form, where
 the diffusion barrier is directly given by $E_o$ 
 if the number of bonds decreases or remains the same
 in the process. In the opposite case
 the energy gain adds to the
 barrier.  Note that the rates  in Eq.\ (\ref{bondcountingrate}) satisfy
 the detailed balance condition  by construction. 
 Apart from the extra barrier, which cancels out in
 (\ref{detailedbalance}), the prescription is equivalent
 with  Metropolis like  rates in equilibrium MC simulations 
 \cite{newman}. 
 
 If only one characteristic $E_o$ is present in the system, the 
 factor $e^{-E_o/(k_B T)}$ appears in all rates and might be taken 
 to re-define the time scale. However, several such barriers may 
 apply to different types of events, e.g. to planar diffusion or
 hops along the border of an existing terrace (step edge diffusion).
 The above discussed Ehrlich-Schwoebel effect, for instance, would
 be represented by a distinct barrier for inter-layer diffusion. 

 Often, several processes are excluded explicitly from the simulation.
 For large enough flux, for instance, the time required for 
 adding an entire monolayer may be much smaller than the typical
 waiting time for a desorption event.  In such a case, desorption
 may be disregarded or considered {\sl forbidden\/}, whereas
 in a vacuum chamber without incoming flux it is clearly
 relevant and the crystal sublimates, see \cite{prbtwosix}
 for the discussion of an example. 

 One  can also expect that the detachment of
 an atom from a terrace edge or from a small island  will
 occur with a very small rate if typical binding energies are
 large.  Hence, one might consider all detachment processes
 {\sl forbidden\/} in an extreme case,
 which reduces the number of possible events considerably. 
 Note that the implementation of irreversible attachment
 immediately violates detailed balance as only one of the rates 
 in Eq.\ (\ref{detailedbalance}) is non-zero. 
 Again, the justification of the simplification hinges 
 on a comparison of the relevant
 time scales which are mainly determined by the incoming flux
 and the substrate temperature.

 The risk of missing relevant microscopic processes 
 is one of the main dangers in KMC simulations, especially in the context
 of material specific investigations.  Complicated
 and -- at first sight -- counter-intuitive concerted moves may 
 be very well relevant in a specific system. This can lead to an
 oversimplified catalogue of events. 
 In addition, wrong rates can be assigned to the implemented 
 events whenever alternative pathways to the same final
 state have been ignored.  

 On the other hand, the potential to explicitly
 {\sl allow\/} or {\sl forbid\/} certain processes is
 one of the major strengths of the KMC approach.
 It allows to investigate the relevance of particular
 microscopic processes and their influence on the macroscopic properties
 of the growing crystal systematically.  
 As just one example for this strategy,
 the role of  step edge diffusion on the morphology and
 scaling behavior of surfaces can be studied by {\sl switching\/}
 the process {\sl on \/} and {\sl off \/} in  simulations,
 see e.g.\ \cite{roleofsed} and references therein.

 \paragraph{\ \\ Implementation of the Monte Carlo Kinetics \ \\}
  The standard textbook realization of the stochastic Monte 
  Carlo kinetics in computer simulations amounts to
  (a) suggesting one of the possible events with equal probability, 
  and (b) evaluate or look up its rate $r_i$ and accept or reject it
      with a corresponding probability. 

  This conceptually simple strategy is easy to implement and has been
  widely used in both equilibrium and off-equilibrium simulations \cite{newman}.
  In terms of efficiency its major drawback is that quite frequently a suggested
  event is rejected and potentially expensive computations have been performed
  without changing the system at all.  

  In contrast, \textit{rejection-free} or \textit{event-based} methods do perform
  one of the possible $n$ events at each step of the procedure. Given a certain
  configuration of the system,  the next process, say event number $i$, 
  has to be selected with the correct probability
  $  {r_i} \left/ {\sum_{i=1}^n \, r_i}. \right. $
  Next, the associated waiting time $t_i$ is evaluated and the physical time in 
  the model is advanced accordingly. For infrequent events as considered here,
  elementary considerations show that the time $t_i$ is exponentially distributed
  with mean value $1/\sum_{k=1}^n r_k$ \cite{krug,newman}.

  The simplest realization of the correct  selection  requires only
  one random number $z \in [0,1] $: the event $i$  which satisfies
  the condition 
  \begin{equation} 
   \textstyle 
   \sum_{k=1}^{i-1} r_k \leq z \sum_{k=1}^n r_k  \leq \sum_{k=1}^i r_k 
   \end{equation}
  is chosen and performed. The costly linear search can be replaced by
  more sophisticated  tree-like representations of the catalogue of events \cite{newman}.
  In  simulations with a limited number of distinct processes
  it is efficient to group the events according to their rate. 
  However, the basic idea of  rejection free simulations remains the same. 

  The price paid for the gain in efficiency is that, at any given time,
  all possible processes must be known and stored. 
  Their number and nature as well as their rates
  can change with every event that occurs in the system.  Depending on the 
  structure of the model this can require sophisticated bookkeeping and
  costly frequent re-evaluations of all rates. 

\section{Conclusion} \label{summary}

  Kinetic Monte Carlo simulations of lattice gas systems
  constitute one of the most widely used tools in the modeling of epitaxial 
  growth. 
  The success of the approach is due to the fact that 
  it is extremely flexible and versatile. 

  Conceptually very simple models can be used in the investigation of basic, 
  perhaps universal properties. It is quite straightforward to set up, 
  for instance, a simple cubic model with only a few distinct
  rates. Such a system can be implemented very efficiently and 
  already allows for the study of various essential phenomena. 

  On the other hand, lattice gas models offer the potential
  for material specific simulations, as well. The success of
  such attempts depends on the availability of detailed information
  about the microscopic processes. To a certain extent, 
  this knowledge can arise from comparison with experiments.
  In turn, simulations allow for  testing hypotheses about the
  microscopic causes of macroscopic surface properties. 

  First principle calculations or approximate theoretical
  treatments are clearly most desirable as a basis of 
  material specific models. A particularly promising route
  seems to be the incorporation of such methods into 
  the on-going simulation.

  A complete description of epitaxial growth requires the
  consideration of processes on many different time and
  length scales. Lattice gas models and 
  Kinetic Monte Carlo simulations in general, will certainly
  play an essential role in the further development of the
  multi-scale approach.

% ------------------------------------------------------------------------

\subsection*{Acknowledgment}
The author  would like to thank the organizers and all participants of the 
MFO Mini-Workshop 
on {\sl Multiscale Modeling in Epitaxial Growth\/}
for the most stimulating atmosphere and many
useful discussions. 
% ------------------------------------------------------------------------

\begin{thebibliography}{12}
\bibitem{krugremark} The author would like to thank J. Krug for  
 drawing his attention to this point.
\end{thebibliography}
\end{document}